\documentclass[twocolumn,showpacs,prd,floatfix,axodraw]{revtex4}
\usepackage{mathrsfs}
\usepackage{graphicx,,booktabs,bm}
\usepackage{overpic}
\usepackage{color}
\usepackage{amssymb}

\usepackage{mathrsfs,bm,amsmath,amssymb}
\usepackage{longtable,lscape}
\usepackage{txfonts}
\usepackage{amssymb}
\usepackage{indentfirst}
\usepackage{graphicx,,booktabs}
\usepackage{multirow,ulem}

\usepackage{color}
\usepackage{amssymb}

\definecolor{cover}{rgb}{0.77,0.87,0.88}
\definecolor{blueone}{rgb}{0.1,0.1,.7}
\definecolor{citec}{rgb}{0.14,0.47,0.09}
\definecolor{two}{rgb}{0.0,0.5,0.}
\definecolor{three}{rgb}{.5,.1,0.15}
\usepackage[bookmarks=true,bookmarksopen=false,plainpages=false,breaklinks=true,
   bookmarksnumbered=true,hypertexnames=false,
   filecolor=blue,urlcolor=three,menucolor=three,
   linkcolor=three,citecolor=blueone, colorlinks,
   anchorcolor=blue,runcolor=pink,frenchlinks=red
   pdfstartview=FitH,pdftitle=title,%
   pdfauthor=author]{hyperref}

\begin{document}
\title{Radiative decay of the $\Xi(1620)$ in a hadronic molecule picture}

\author{HongQiang Zhu}
\affiliation{College of Physics and Electronic Engineering, Chongqing Normal University,  Chongqing 401331,China}

\author{Feng Yang}
\affiliation{School of Physical Science and Technology, Southwest Jiaotong University, Chengdu 610031,China}

\author{Yin Huang \footnote{corresponding author}}
\email{huangy2019@swjtu.edu.cn}
\affiliation{School of Physical Science and Technology, Southwest Jiaotong University, Chengdu 610031,China}

\date{\today}
\begin{abstract}
Last year, the $\Xi(1620)$ state that is cataloged in the Particle Data Group
(PDG) with only one star is reported again in the $\Xi^{-}\pi^{+}$ final state by
the Belle Collaboration.  Its properties not only the spectroscopy but also the decay width
cannot be simply explained in the context of conventional constituent quark models.
This intrigues an active discussion on the structure of this resonance.
In this work, we study the radiative decays of the newly observed $\Xi(1620)$ assuming that
it is a meson-baryon molecular state of $\Lambda\bar{K}$ and $\Sigma\bar{K}$ with spin-parity
$J^P=1/2^{-}$ in our previous work.  The partial decay widths of the $\Lambda\bar{K}-\Sigma\bar{K}$
molecular state into $\Xi\gamma$ and $\Xi\pi\gamma$ final states through hadronic loops are
evaluated with the help of the effective Lagrangians. The partial widths for the $\Xi(1620)^0\to\gamma\Xi$
and $\Xi(1620)^0\to\gamma\Xi\pi$ are evaluated to be about $118.76-174.21$ KeV and $58.19-68.75$ eV,
respectively, which may be accessible for the LHCb.  If the $\Xi(1620)$ is $\Lambda\bar{K}-\Sigma\bar{K}$
molecule, the radiative transition strength $\Xi(1620)^0\to\gamma\bar{K}\Lambda$ is quite small and
the decay width is of the order of 0.01 eV.   Future experimental measurements of these processes can be
useful to test the molecule interpretations of the $\Xi(1620)$.
\end{abstract}


\maketitle
\section{INTRODUCTION}
Searching for hadrons beyond the quark model becomes one of the most important topics in the community
of hadron physics.  In the conventional quark model, a hadron is composed of $q\bar{q}$ as a meson or $qqq$
as a baryon.  It is natural to expect the existence of hadrons composed of more quarks, which are called
exotic states.  Due to the great  experimental progress in the past 20 years, more and more exotic states
have been observed~\cite{Zyla:2020zbs}.   Last year, the $\Xi(1620)$ state that is cataloged in the Particle
Data Group(PDG)~\cite{Zyla:2020zbs} with only one star is reported again in the $\Xi^{-}\pi^{+}$ final state
by the Belle Collaboration~\cite{Sumihama:2018moz}.  The observed resonance parameters of the structure are
\begin{align}
&M=1610.4\pm{6.0}(stat)^{+5.9}_{-3.5}(syst)~~ {\rm MeV},\nonumber\\
&\Gamma=59.9\pm4.8(stat)^{+2.8}_{-3.0}(syst)~~{\rm MeV},
\end{align}
which are consistent with the earlier measured values~\cite{Ross:1972bf,Briefel:1977bp}.
Its spin-parity, however, remains undetermined.

From the observed decay model, the $\Xi(1620)$ is a conventional baryon composed of $uss$ or $dss$.  However,
its properties not only the spectroscopy but also the decay width cannot be simply explained in the context of
conventional constituent quark models~\cite{Capstick:1986bm,Blask:1990ez,Azimov:2003bb}.
As indicated in Refs.~\cite{Ramos:2002xh,GarciaRecio:2003ks,Gamermann:2011mq,Miyahara:2016yyh,Oh:2007cr,Wang:2019krq,Chen:2019uvv,Huang:2020taj},
the $\Xi(1620)$ can be understood as a molecular state in comparison with the Belle data~\cite{Sumihama:2018moz}.
A molecular state with a narrow width and a mass around 1606 MeV was predicted in the unitarized coupled channels
approach~\cite{Ramos:2002xh,GarciaRecio:2003ks,Gamermann:2011mq,Miyahara:2016yyh}.  However, a $\Xi$ bound state with
a mass about 1620 MeV and $J^P=1/2^{-}$ is predicted by studying the $\bar{K}\Lambda$ interaction in the framework
of the One-Boson-Exchange (OBE) model~\cite{Chen:2019uvv}.  In other words, the Ref.~\cite{Chen:2019uvv} consider
$\Xi(1620)$ as a pure molecular state composes of $\bar{K}\Lambda$ component.  A possible explanation for these results
of Refs.~\cite{Ramos:2002xh,GarciaRecio:2003ks,Gamermann:2011mq,Miyahara:2016yyh,Chen:2019uvv} is that the $\Xi$(1620)
has a larger $\bar{K}\Lambda$ component~\cite{Wang:2019krq,Huang:2020taj}.   Moreover, the $\bar{K}\Sigma$ component
can not be underestimated to reproduce the total decay width of the $\Xi(1620)$ and  a study of the spectroscopy alone
does not give a complete picture of its nature~\cite{Huang:2020taj}.

From the above discussion, the $\Xi(1620)$ may be a molecular state. However, at present, we cannot fully exclude other possible
explanations such as a mixture of three quark and five quark components (as long as quantum numbers allow, it might well be the case).
More efforts are necessary to distinguish whether it is a molecular state or a compact multi-quark state.  The photon coupling with quark
is really different from the coupling of the photon to the constituent $\bar{K}\Lambda$ and $\bar{K}\Sigma$  of the $\Xi(1620)$~\cite{Koniuk:1979vy}.
Hence, a precise measurement of the radiative decays is useful to test different interpretations of $\Xi(1620)$.   In this work,  we study
the radiative decay of the $\Xi(1620)$ in the hadronic molecule approach developed in our previous study~\cite{Huang:2020taj}.

This  paper  is  organized  as  follows.  The  theoretical
formalism is explained in Sec.II. The predicted partial decay  widths  are  presented in Sec.III,  followed  by  a  short
summary in the last section.

\section{Theoretical FORMALISM}
\begin{figure*}[htbp]
\begin{center}
\includegraphics[scale=0.60]{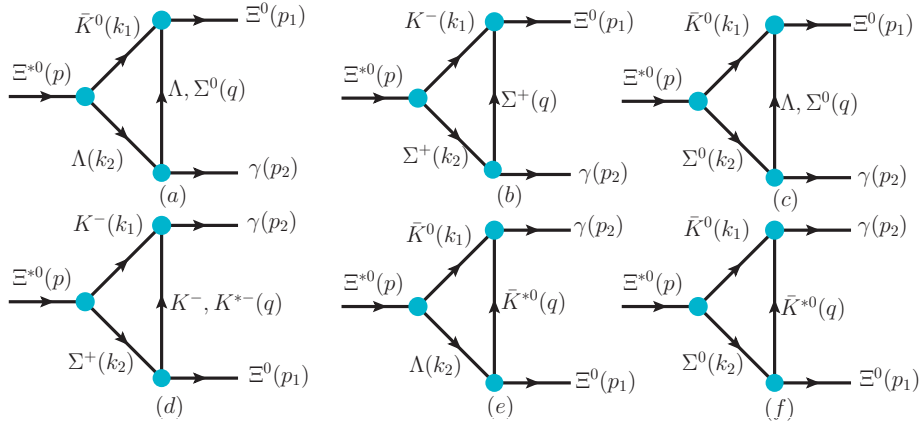}
\caption{Feynman diagrams for the $\Xi^{*0}\to{}\gamma\Xi^{0}$ decay processes.  We also show
the definitions of the kinematics ($p,k_1,k_2,p_1,p_2$, and $q$) used in the calculation.}\label{fety}
\end{center}
\end{figure*}

\begin{figure}[htbp]
\centering
\includegraphics[scale=0.6]{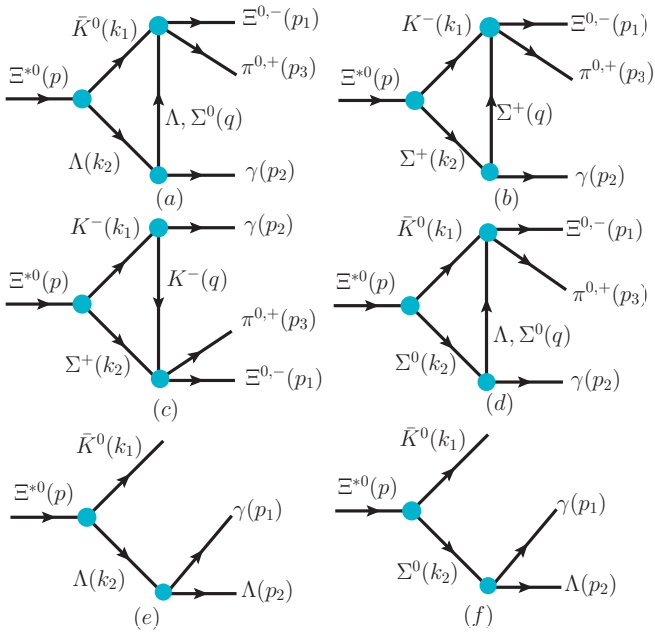}
\caption{Feynman diagrams for the $\Xi^{*0}\to{}\gamma\Xi^{0}\pi^0$,$\gamma\Xi^{-}\pi^{+}$, and $\gamma\bar{K}\Lambda$ decay processes.  We also show
the definitions of the kinematics ($p,k_1,k_2,p_1,p_2$,$p_3$, and $q$) used in the calculation.} 
\label{fig:example1}
\end{figure}

In our previous study~\cite{Huang:2020taj}, the total decay width of $\Xi(1620)$ can be reproduced with the assumption that the
$\Xi(1620)$ is an $S-$wave $\bar{K}\Lambda-\bar{K}\Sigma$ bound state with $J^P=1/2^{-}$.  According to the molecular scenario,
the radiative decay widths $\Xi(1620)\to\gamma\Xi$, $\Xi(1620)\to\gamma\bar{K}\Lambda$ and, $\Xi(1620)\to\gamma\pi\Xi$ are
studied to elucidate the internal structure of the $\Xi(1620)$.  In order to calculate the radiation decay, we first employ
the weinberg compositeness rule to determine $\Xi(1620)$
couplings to hers constituents $\bar{K}Y$ ($Y\equiv\Lambda,\Sigma$). The radiative decays then follow from the exchange of a suitable
hadron between the $\bar{K}Y$ pair, which then transforms into $\Xi\gamma$,$\gamma\bar{K}\Lambda$, and $\Xi\pi\gamma$.
The corresponding Feynman diagrams are shown in Fig.~(\ref{fety}) and Fig.~(\ref{fig:example1}).

To compute the diagrams shown in Figs.~(\ref{fety}-\ref{fig:example1}), we need the effective Lagrangian densities for the relevant interaction vertices.
For the $\Xi(1620)\bar{K}Y$ vertices, the Lagrangian densities can be written as~\cite{Huang:2020taj,Huang:2018bed,Dong:2010gu}
\begin{align}
{\cal{L}}_{\Xi(1620)}(x)&=g_{\Xi(1620)\bar{K}Y}\int{}d^4y\Phi(y^2)\bar{K}(x+\omega_{Y}y)\nonumber\\
                                  &\times{}Y(x-\omega_{\bar{K}}y)\bar{\Xi}(1620)(x)\label{eq1}.
\end{align}
Where $\omega_{\bar{K}}=m_{\bar{K}}/(m_{\bar{K}}+m_{Y})$ and $\omega_{Y}=m_{Y}/(m_{\bar{K}}+m_{Y})$.
For an isovector baryon $\Sigma$, $Y$ should be replaced with $\vec{Y}\cdot\vec{\tau}$, where $\tau$ is
the isospin matrix.  The $\Phi(y^2)$ is an effective correlation function that is introduced to describe
the distribution of the constituents,  $\bar{K}$
and $Y$, in the hadronic molecular $\Xi(1620)$ state, which is often chosen to be of the  following form ~\cite{Huang:2020taj,Huang:2018bed,Dong:2010gu,Huang:2019qmw,Faessler:2007us,Dong:2009yp,Dong:2009uf,Dong:2017gaw,Xiao:2019mst,Huang:2018wgr} and references therein,
\begin{align}
\Phi(p_E^2)\doteq\exp(-p_E^2/\beta^2)
\end{align}
with $p_E$ being the Euclidean Jacobi momentum and $\beta$ being the size parameter which characterizes the distribution
of the components inside the molecule.  At present, the value of $\beta=1.0$ is determined by experimental data~\cite{Huang:2020taj,Huang:2018bed,Dong:2010gu,Huang:2019qmw,Faessler:2007us,Dong:2009yp,Dong:2009uf,Dong:2017gaw,Xiao:2019mst,Huang:2018wgr}(and references therein).

\begin{figure*}[htbp]
\begin{center}
\includegraphics[scale=0.65]{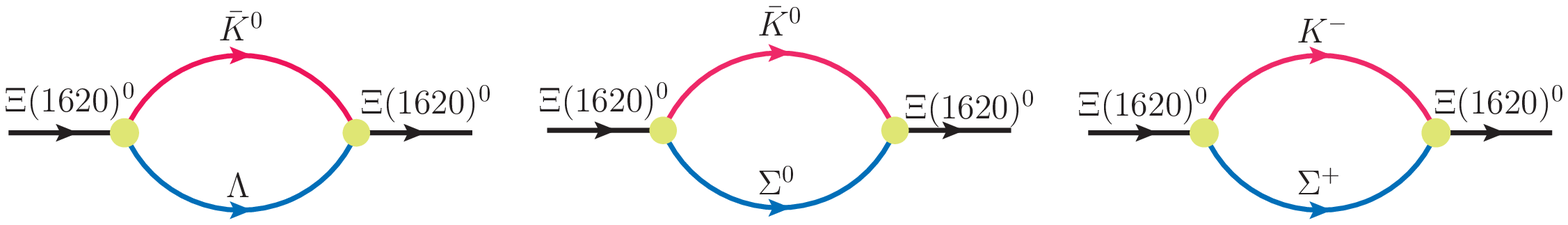}
\end{center}
\caption{Self-energy of the $\Xi$(1620) state.}\label{msef}
\end{figure*}

The coupling constant $g_{\Xi(1620)\bar{K}Y}$ is determined by the compositeness condition~\cite{Huang:2020taj,Huang:2018bed,Dong:2010gu,Huang:2019qmw,Faessler:2007us,Dong:2009yp,Dong:2009uf,Dong:2017gaw,Xiao:2019mst,Huang:2018wgr},
which implies that the renormalization constant of the bound state wave function $\Xi(1620)$ is set to zero
\begin{align}
Z_{\Xi(1620)}=x_{\bar{K}\Sigma}+x_{\bar{K}\Lambda}-\frac{d\Sigma_{\Xi(1620)}}{dk\!\!\!/_0}|_{k\!\!\!/_0=m_{\Xi(1620)}}=0\label{eqn3}.
\end{align}
Where $x_{AB}$ is the probability to find the $\Xi(1620)$ in the hadronic state $AB$ with normalization $x_{\bar{K}\Sigma}+x_{\bar{K}\Lambda}=1.0$.
The $\Sigma_{\Xi(1620)}$ is self-energy of the $\Xi(1620)$ and can be computed follow the Feynmann diagrams shown in Fig.(\ref{msef})

\begin{align}
\Sigma_{\Xi(1620)}(k_0)&=\sum_{Y=\Lambda,\Sigma^0,\Sigma^{+}}({\cal{C}}_{Y})^2g^2_{\Xi(1620)\bar{K}Y}\int_0^{\infty}d\alpha\int_0^{\infty}d\eta\nonumber\\
                       &\times\frac{(-\frac{\Delta}{2z}k\!\!\!/_0+m_Y)}{16\pi^2z^2}\exp\{-\frac{1}{\beta^2}[(-2\omega_Y^2-\eta\nonumber\\
                       &+\frac{\Delta^2}{4z})k_0^2+\alpha{}m_Y^2+\eta{}m^2_{\bar{K}}]\}\label{eqn1}.
\end{align}
where the $z=2+\alpha+\eta$ and $\Delta=-4\omega_Y-2\eta$.  The $k_0^2=m^2_{\Xi(1620)}$ with $k_0, m_{\Xi(1620)}$ denoting the four momenta
and the mass of the $\Xi(1620)$, respectively,
$k_1$, $m_{\bar{K}}$, and $m_{Y}$ are the four-momenta, the mass of the $\bar{K}$ meson, and the mass of the $Y$ baryon, respectively.
Here, we set $m_{\Xi(1620)}=m_{Y}+m_{\bar{K}}-E_b$ with $E_b$ being the binding energy of $\Xi(1620)$.  Isospin symmetry implies that
\begin{align}
{\cal{C}}_Y=\left\{
\begin{aligned}
1           &  & {Y=\Lambda} \\
\sqrt{1/3}         &  & {Y=\Sigma^0} \\
-\sqrt{2/3}         &  & {Y=\Sigma^+}.
\end{aligned}
\right.
\end{align}

To estimate the radiative decays of the diagrams shown in Figs.~(\ref{fety}-\ref{fig:example1}), we need the effective Lagrangian
densities related to the photon fields, which are~\cite{Kim:2012pz,Kim:2018qfu}
\begin{align}
{\cal{L}}_{\gamma\Sigma\Sigma}&=-\bar{\Sigma}[e_{\Sigma}A\!\!\!/-\frac{e{\kappa_{\Sigma}}}{2m_N}\sigma_{\mu\nu}\partial^{\nu}A^{\mu}]\Sigma,\\
{\cal{L}}_{\gamma\Lambda\Lambda}&=\frac{e\kappa_{\Lambda}}{2m_{N}}\bar{\Lambda}\sigma_{\mu\nu}\partial^{\nu}A^{\mu}\Lambda,\\
{\cal{L}}_{\gamma\Sigma\Lambda}&=\frac{e\mu_{\Sigma\Lambda}}{2m_{N}}\bar{\Sigma}^0\sigma_{\mu\nu}\partial^{\nu}A^{\mu}\Lambda,\\
{\cal{L}}_{K^{*}K\gamma}&=\frac{g_{K^{*+}K^{+}\gamma}}{4}e\epsilon^{\mu\nu\alpha\beta}F_{\mu\nu}K^{*+}_{\alpha\beta}K^{-}\nonumber\\
                        &+\frac{g_{K^{*0}K^{0}\gamma}}{4}e\epsilon^{\mu\nu\alpha\beta}F_{\mu\nu}K^{*0}_{\alpha\beta}\bar{K}^{0}+h.c.,\label{eqw1}\\
{\cal{L}}_{KK\gamma}&=ieA_{\mu}K^{-} \overleftrightarrow{\partial}^{\mu}K^{+}\label{eqw2}.
\end{align}
Where the strength tensor are defined as $\sigma_{\mu\nu}=\frac{i}{2}(\gamma_{\mu}\gamma_{\nu}-\gamma_{\nu}\gamma_{\mu})$, $F_{\mu\nu}=\partial_{\mu}A_{\nu}-\partial_{\nu}A_{\mu}$, and $K^{*}_{\mu\nu}=\partial_{\mu}K^{*}_{\nu}-\partial_{\nu}K^{*}_{\mu}$.
The $M_N$ is the mass of the $p$ and $\alpha=e^2/4\pi=1/137$ is the electromagnetic fine structure constant.
The anomalous and transition magnetic moments of the baryons are given by the PDG~\cite{Zyla:2020zbs}
and are shown in Tab.(\ref{tabke1})
\begin{table}[h!]
\centering
\caption{The anomalous and transition magnetic moments. }\label{tabke1}
\begin{tabular}{ccc}
\hline \hline
  ~~~~~~~~~~$\kappa_{\Sigma^{-}}=-0.16$  & ~~~~~~~~~~$\kappa_{\Sigma^{0}}=0.65$     &~~~~~~~~~~$\kappa_{\Sigma^{+}}=1.46$~~~~~\\
  ~~~~~~~~~~$\kappa_{\Lambda}=-0.61$     & ~~~~~~~~~~$\mu_{\Sigma\Lambda}=1.61$                                           \\
\hline\hline
\end{tabular}
\end{table}

The coupling constant $g_{K^{*+}K^{+}\gamma}$ and $g_{K^{*0}K^{0}\gamma}$ are introduced to
get consistent results with experimental measurements of $K^{*+}\to{}K^{+}\gamma$ and $K^{*0}\to{}K^{0}\gamma$.
The theoretical decay widths of $K^{*+}\to{}K^{+}\gamma$ and $K^{*0}\to{}K^{0}\gamma$ are
\begin{align}
&\Gamma(K^{*+}\to{}K^{+}\gamma)=\frac{\alpha{}g^2_{K^{*+}K^{+}\gamma}}{24}m_{K^{*+}}(m^2_{K^{*+}}-m^2_{K^{+}}),\\
&\Gamma(K^{*0}\to{}K^{0}\gamma)=\frac{\alpha{}g^2_{K^{*0}K^{0}\gamma}}{24}m_{K^{*0}}(m^2_{K^{*0}}-m^2_{K^{0}}).
\end{align}
According to the experimental widths $\Gamma(K^{*+}\to{}K^{+}\gamma)=0.0503$ keV~\cite{Zyla:2020zbs} and $\Gamma(K^{*0}\to{}K^{0}\gamma)=0.125$ keV~\cite{Zyla:2020zbs}, the coupling constant $g_{K^{*}K\gamma}$ is fixed as
\begin{align}
g_{K^{*+}K^{+}\gamma}=0.580~ {\rm GeV^{-1}},~~~~~g_{K^{*0}K^{0}\gamma}=-0.904~ {\rm GeV^{-1}},
\end{align}
and the signs of these coupling constants are fixed by the quark model.

In addition to the Lagrangians above, the meson-baryon interactions are also needed and can be obtained from the following chiral Lagrangians~\cite{Garzon:2012np,Oset:1997it}
\begin{align}
{\cal{L}}_{VBB}&=g(\langle\bar{B}\gamma_{\mu}[V^{\mu},B]\rangle+\langle\bar{B}\gamma_{\mu}B\rangle\langle{}V^{\mu}\rangle)\label{vbb},\\
{\cal{L}}_{PBB}&=\frac{F}{2}\langle\bar{B}\gamma_{\mu}\gamma_5[u^{\mu},B]\rangle+\frac{D}{2}\langle\bar{B}\gamma_{\mu}\gamma_5\{u^{\mu},B\}\rangle\label{Pbb},\\
{\cal{L}}_{PBPB}&=\frac{i}{4f^2}\langle\bar{B}\gamma^{\mu}[(P\partial_{\mu}P-\partial_{\mu}PP)B\nonumber\\
                &-B(P\partial_{\mu}P-\partial_{\mu}PP)]\rangle
\end{align}
where $g=4.64$, $F=0.51$, $D=0.75$~\cite{Huang:2020taj,Garzon:2012np,Borasoy:1998pe} and at the lowest order $u^{\mu}=-\sqrt{2}\partial^{\mu}P/f$
with $f=93$ MeV, and $\langle...\rangle$ denotes trace in the flavor space.  The $B$,$P$, and $V^{\mu}$ are the $SU(3)$
pseudoscalar meson, vector meson, and baryon octet matrices, respectively, which are
\begin{equation}
B=
\left(
  \begin{array}{ccc}
    \frac{1}{\sqrt{2}}\Sigma^{0}+\frac{1}{\sqrt{6}}\Lambda  & \Sigma^{+}                                               &  p      \\
     \Sigma^{-}                                             & -\frac{1}{\sqrt{2}}\Sigma^{0}+\frac{1}{\sqrt{6}}\Lambda  &  n       \\
     \Xi^{-}                                                & \Xi^{0}                                                  &  -\frac{2}{\sqrt{6}}\Lambda     \\
  \end{array}
\right),
\end{equation}
\begin{equation}
P=
\left(
  \begin{array}{ccc}
    \frac{\pi^{0}}{\sqrt{2}}+\frac{\eta}{\sqrt{6}} &    \pi^{+}                                        &       K^{+}\\
    \pi^{-}                                       &    -\frac{\pi^{0}}{\sqrt{2}}+\frac{\eta}{\sqrt{6}} &       K^{0}\\
    K^{-}                                         &    \bar{K}^{0}                                     &       -\frac{2}{\sqrt{6}}\eta
  \end{array}
\right),\label{eq7}
\end{equation}
\begin{equation}
V_{\mu}=
\left(
  \begin{array}{cccc}
    \frac{1}{\sqrt{2}}(\rho^{0}+\omega) & \rho^{+}                             &  K^{*+}      \\
    \rho^{-}                            & \frac{1}{\sqrt{2}}(-\rho^{0}+\omega) &  K^{*0}       \\
     K^{*-}                             & \bar{K}^{*0}                         &  \phi       \\
  \end{array}
\right)_{\mu}.
\end{equation}

Putting all the pieces together, we obtain the following decay amplitudes,
\begin{align}
{\cal{M}}_{a}&(\Xi^{*0}\to\Xi^{0}\gamma)=i(i)^3\{\frac{(D-3F)\kappa_{\Lambda}}{2\sqrt{3}},\frac{(F+D)\mu_{\Sigma\Lambda}}{2}\}\nonumber\\
                   &\times{}\frac{eg_{\Xi^{*}\Lambda\bar{K}}}{4m_Nf}\int\frac{d^4q}{(2\pi)^4}\Phi[(k_1\omega_{\Lambda}-k_2\omega_{\bar{K}^{0}})^2]\nonumber\\
                   &\times\bar{u}(p_1)k\!\!\!/_1\gamma_5\frac{q\!\!\!/+m_{Y}}{q^2-m^2_{Y}}(\gamma_{\mu}p\!\!\!/_2-p\!\!\!/_2\gamma_{\mu})\frac{k\!\!\!/_2+m_{\Lambda}}{k_2^2-m^2_{\Lambda}}\nonumber\\
                   &\times{}u(p)\frac{1}{k_1^2-m^2_{\bar{K}^{0}}}\epsilon^{*\mu}(p_2),\\
{\cal{M}}_{b}&(\Xi^{*0}\to\Xi^{0}\gamma)=-i(i)^3e\frac{D+F}{\sqrt{2}f}C_{Y}g_{\Xi^{*}\Sigma^+\bar{K}^-}\int\frac{d^4q}{(2\pi)^4}\nonumber\\
                   &\times{}\Phi[(k_1\omega_{\Sigma^+}-k_2\omega_{\bar{K}^{-}})^2]\bar{u}(p_1)k\!\!\!/_1\gamma_5\nonumber\\
                   &\times\frac{q\!\!\!/+m_{Y}}{q^2-m^2_{Y}}[\gamma_{\mu}+\frac{\kappa_{Y}}{4m_N}(\gamma_{\mu}p\!\!\!/_2-p\!\!\!/_2\gamma_{\mu})]\nonumber\\
                   &\times\frac{k\!\!\!/_2+m_{\Sigma^{+}}}{k_2^2-m^2_{\Sigma^{+}}}u(p)\frac{1}{k_1^2-m^2_{K^{-}}}\epsilon^{*\mu}(p_2),
\end{align}
\begin{align}
{\cal{M}}_{c}&(\Xi^{*0}\to\Xi^{0}\gamma)=i(i)^3\{\frac{(D-3F)\mu_{\Sigma\Lambda}}{2\sqrt{3}},\frac{(F+D)\kappa_{\Sigma^0}}{2}\}\nonumber\\
                   &\times{}\frac{eg_{\Xi^{*}\Lambda\bar{K}}C_{Y}}{4m_Nf}\int\frac{d^4q}{(2\pi)^4}\Phi[(k_1\omega_{\Sigma^0}-k_2\omega_{\bar{K}^{0}})^2]\nonumber\\
                   &\times\bar{u}(p_1)k\!\!\!/_1\gamma_5\frac{q\!\!\!/+m_{Y}}{q^2-m^2_{Y}}(\gamma_{\mu}p\!\!\!/_2-p\!\!\!/_2\gamma_{\mu})\frac{k\!\!\!/_2+m_{\Sigma^0}}{k_2^2-m^2_{\Sigma^0}}\nonumber\\
                   &\times{}u(p)\frac{1}{k_1^2-m^2_{\bar{K}^{0}}}\epsilon^{*\mu}(p_2),\\
{\cal{M}}^{K}_{d}&(\Xi^{*0}\to\Xi^{0}\gamma)=-i(i)^3\frac{D+F}{\sqrt{2}f}eC_Yg_{\Xi^{*}\Sigma^{+}K^-}\nonumber\\
                   &\times{}\int\frac{d^4q}{(2\pi)^4}\Phi[(k_1\omega_{\Sigma^{+}}-k_2\omega_{{K}^{-}})^2]\bar{u}(p_1)q\!\!\!/\gamma_5\nonumber\\
                   &\times{}\frac{k\!\!\!/_2+m_{\Sigma^{+}}}{k_2^2-m^2_{\Sigma^{+}}}u(p)\frac{1}{k_1^2-m^2_{K^-}}\frac{1}{q^2-m^2_{K^+}}\nonumber\\
                   &\times(q-k_1)\cdot\epsilon^{*}(p_2),\\
{\cal{M}}^{K^{*}}_{d}&(\Xi^{*0}\to\Xi^{0}\gamma)=(i)^3\frac{egC_Yg_{K^{*}K\gamma}}{4}\int\frac{d^4q}{(2\pi)^4}\nonumber\\
                   &\times{}\Phi[(k_1\omega_{\Sigma^{+}}-k_2\omega_{{K}^{-}})^2]\bar{u}(p_1)\gamma_{\rho}\frac{k\!\!\!/_2+m_{\Sigma^{+}}}{k_2^2-m^2_{\Sigma^{+}}}u(p)\nonumber\\
                   &\times{}\frac{1}{k_1^2-m^2_{K^-}}\frac{-g^{\rho\sigma}+q^{\rho}q^{\sigma}/m^2_{K^{*-}}}{q^2-m^2_{K^{*-}}}\epsilon^{\mu\nu\alpha\beta}\nonumber\\
                   &\times{}(p_{2\mu}g_{\nu\eta}-p_{2\nu}g_{\mu\eta})(q_{\alpha}g_{\beta\sigma}-q_{\beta}g_{\alpha\sigma})\epsilon^{*\eta}(p_2),\\
{\cal{M}}_{e}&(\Xi^{*0}\to\Xi^{0}\gamma)=(i)^3\frac{\sqrt{6}egC_Yg_{K^{*}K\gamma}}{8}\int\frac{d^4q}{(2\pi)^4}\nonumber\\
                   &\times{}\Phi[(k_1\omega_{\Lambda}-k_2\omega_{\bar{K}^{0}})^2]\bar{u}(p_1)\gamma_{\rho}\frac{k\!\!\!/_2+m_{\Lambda}}{k_2^2-m^2_{\Lambda}}u(p)\nonumber\\
                   &\times{}\frac{1}{k_1^2-m^2_{\bar{K}^0}}\frac{-g^{\rho\sigma}+q^{\rho}q^{\sigma}/m^2_{\bar{K}^{*0}}}{q^2-m^2_{\bar{K}^{*0}}}\epsilon^{\mu\nu\alpha\beta}\nonumber\\
                   &\times{}(p_{2\mu}g_{\nu\eta}-p_{2\nu}g_{\mu\eta})(q_{\alpha}g_{\beta\sigma}-q_{\beta}g_{\alpha\sigma})\epsilon^{*\eta}(p_2),\\
{\cal{M}}_{f}&(\Xi^{*0}\to\Xi^{0}\gamma)=-(i)^3\frac{egC_Yg_{K^{*}K\gamma}}{4\sqrt{2}}\int\frac{d^4q}{(2\pi)^4}\nonumber\\
                   &\times{}\Phi[(k_1\omega_{\Sigma^{0}}-k_2\omega_{\bar{K}^{0}})^2]\bar{u}(p_1)\gamma_{\rho}\frac{k\!\!\!/_2+m_{\Sigma^{0}}}{k_2^2-m^2_{\Sigma^{0}}}u(p)\nonumber\\
                   &\times{}\frac{1}{k_1^2-m^2_{\bar{K}^0}}\frac{-g^{\rho\sigma}+q^{\rho}q^{\sigma}/m^2_{\bar{K}^{*0}}}{q^2-m^2_{\bar{K}^{*0}}}\epsilon^{\mu\nu\alpha\beta}\nonumber\\
                   &\times{}(p_{2\mu}g_{\nu\eta}-p_{2\nu}g_{\mu\eta})(q_{\alpha}g_{\beta\sigma}-q_{\beta}g_{\alpha\sigma})\epsilon^{*\eta}(p_2),
\end{align}
where $\{{\cal{A}}$, and ${\cal{B}}\}$ are $\Lambda$ and $\Sigma$ baryon exchange,respectively.   The amplitudes of the $\Xi^{0}(1620)\to\gamma\pi\Xi$ can be also easy obtained
\begin{align}
\begin{split}
{\cal{M}}_{a}&(\Xi^{*0}\to\pi^0\Xi^{0}\gamma,\pi^{+}\Xi^{-}\gamma)= i^3\frac{eC_{Y}g_{\Xi^{*}\bar{K}^0\Lambda}}{32m_{N}f^2}\left \{
\begin{array}{ll}
   \{-\sqrt{3}\kappa_{\Lambda},\mu_{\Sigma\Lambda}\}\\
    \{\sqrt{6}\kappa_{\Lambda},\sqrt{2}\mu_{\Sigma\Lambda}\} \\
\end{array}
\right.
\end{split}
\nonumber\\
&\times{}\int\frac{d^4q}{(2\pi)^4}\Phi[(k_1\omega_{\Lambda}-k_2\omega_{\bar{K}^{0}})^2]\bar{u}(p_1)k\!\!\!/_1\frac{q\!\!\!/+m_{Y}}{q^2+m^2_{Y}}\nonumber\\
&\times{}(\gamma_{\mu}p\!\!\!/_2-p\!\!\!/_2\gamma_{\mu})\frac{k\!\!\!/_2+m_{\Lambda}}{k_2^2-m^2_{\Lambda}}u(p)\epsilon^{*\mu}\frac{1}{k_1^2-m^2_{\bar{K}^0}},\\
\begin{split}
{\cal{M}}_{b}&(\Xi^{*0}\to\pi^0\Xi^{0}\gamma,\pi^{+}\Xi^{-}\gamma)= i^3\frac{eC_{Y}g_{\Xi^{*}K^{-}\Sigma^{+}}}{4f^2}\left \{
\begin{array}{ll}
   \frac{1}{\sqrt{2}}\\
    1 \\
\end{array}
\right.
\end{split}
\nonumber
\end{align}

\begin{align}
&\times{}\int\frac{d^4q}{(2\pi)^4}\Phi[(k_1\omega_{\Sigma^{+}}-k_2\omega_{K^{-}})^2]\bar{u}(p_1)k\!\!\!/_1\frac{q\!\!\!/+m_{Y}}{q^2+m^2_{Y}}\nonumber\\
&\times{}[\gamma_{\mu}+\frac{\kappa_{\Sigma^{+}}}{4m_N}(\gamma_{\mu}p\!\!\!/_2-p\!\!\!/_2\gamma_{\mu})]\frac{k\!\!\!/_2+m_{\Sigma^{+}}}{k_2^2-m^2_{\Sigma^{+}}}u(p)\epsilon^{*\mu}\frac{1}{k_1^2-m^2_{K^-}},\\
\begin{split}
{\cal{M}}_{c}&(\Xi^{*0}\to\pi^0\Xi^{0}\gamma,\pi^{+}\Xi^{-}\gamma)= i^3\frac{eC_{Y}g_{\Xi^{*}K^{-}\Sigma^{+}}}{4f^2}\left \{
\begin{array}{ll}
   \frac{1}{\sqrt{2}}\\
    1 \\
\end{array}
\right.
\end{split}
\nonumber\\
&\times{}\int\frac{d^4q}{(2\pi)^4}\Phi[(k_1\omega_{\Sigma^{+}}-k_2\omega_{K^{-}})^2]\bar{u}(p_1)q\!\!\!/\frac{k\!\!\!/_2+m_{\Sigma^+}}{k_2^2+m^2_{\Sigma^{+}}}u(p)\nonumber\\
&\times{}(q^{\mu}+k^{\mu}_{1})\frac{1}{k_1^2-m^2_{K^{-}}}\frac{1}{q^2-m^2_{K^{-}}}\epsilon^{*}_{\mu},\\
\begin{split}
{\cal{M}}_{d}&(\Xi^{*0}\to\pi^0\Xi^{0}\gamma,\pi^{+}\Xi^{-}\gamma)= i^3\frac{eC_{Y}g_{\Xi^{*}\bar{K}^0\Sigma^0}}{32m_{N}f^2}\left \{
\begin{array}{ll}
   \{-\sqrt{3}\mu_{\Sigma\Lambda},\kappa_{\Sigma^0}\}\\
    \{\sqrt{6}\mu_{\Sigma\Lambda},\sqrt{2}\kappa_{\Sigma^0}\} \\
\end{array}
\right.
\end{split}
\nonumber\\
&\times{}\int\frac{d^4q}{(2\pi)^4}\Phi[(k_1\omega_{\Sigma^0}-k_2\omega_{\bar{K}^{0}})^2]\bar{u}(p_1)k\!\!\!/_1\frac{q\!\!\!/+m_{Y}}{q^2+m^2_{Y}}\nonumber\\
&\times{}(\gamma_{\mu}p\!\!\!/_2-p\!\!\!/_2\gamma_{\mu})\frac{k\!\!\!/_2+m_{\Sigma^0}}{k_2^2-m^2_{\Sigma^0}}u(p)\epsilon^{*\mu}\frac{1}{k_1^2-m^2_{\bar{K}^0}},\\
{\cal{M}}_{e}&(\Xi^{*0}\to\bar{K}^0\Lambda\gamma)=-i\frac{e{\kappa_{\Lambda}}g_{\Xi^{*}\bar{K}^0\Lambda}C_Y}{4m_N}\Phi[(k_1\omega_{\Lambda}-k_2\omega_{\bar{K}^{0}})^2]\nonumber\\
             &\times\bar{u}(p_2)(\gamma^{\mu}p\!\!\!/_1-p\!\!\!/_1\gamma^{\mu})\frac{k\!\!\!/_2+m_{\Lambda}}{k_2^2-m^2_{\Lambda}}u(p)\epsilon^{*}_{\mu}(p_1),\\
{\cal{M}}_{f}&(\Xi^{*0}\to\bar{K}^0\Lambda\gamma)=-i\frac{e\mu_{\Sigma\Lambda}g_{\Xi^{*}\bar{K}^0\Sigma^0}C_Y}{4m_N}\Phi[(k_1\omega_{\Sigma^0}-k_2\omega_{\bar{K}^{0}})^2]\nonumber\\
             &\times\bar{u}(p_2)(\gamma^{\mu}p\!\!\!/_1-p\!\!\!/_1\gamma^{\mu})\frac{k\!\!\!/_2+m_{\Sigma^0}}{k_2^2-m^2_{\Sigma^0}}u(p)\epsilon^{*}_{\mu}(p_1)
\end{align}
where the expressions in the curly brackets, $\left \{
\begin{array}{ll}
   A\\
    B \\
\end{array}
\right.$, are for the $\Xi^{*0}\to\pi^0\Xi^{0}\gamma$ and $\Xi^{*0}\to\pi^{+}\Xi^{-}\gamma$, respectively.

With above, the total amplitudes of the $\Xi^{*0}\to\Xi^{0}\gamma$, $\Xi^{*0}\to\Xi^{0}\pi^0\gamma$, and
$\Xi^{*0}\to\pi^{+}\Xi^{-}\gamma$ are the sum of these individual amplitudes, respectively,
\begin{align}
&{\cal{M}}^{T}(\Xi^{*0}\to\Xi\gamma)=\sum_{i=a,b,c,d,e,f}{\cal{M}}_{i}(\Xi^{*0}\to\Xi\gamma),\\
&{\cal{M}}^{T}(\Xi^{*0}\to\pi\Xi\gamma)=\sum_{i=a,b,c,d}{\cal{M}}_{i}(\Xi^{*0}\to\pi\Xi\gamma),\\
&{\cal{M}}^{T}(\Xi^{*0}\to\bar{K}\Lambda\gamma)=\sum_{i=e.f}{\cal{M}}_{i}(\Xi^{*0}\to\bar{K}\Lambda\gamma).
\end{align}

Once the amplitudes are determined, the corresponding partial decay width can be easily obtained, which reads as,
\begin{align}
d\Gamma(\Xi(1620)^{0}&\to\gamma\Xi)=\frac{1}{2J+1}\frac{1}{32\pi^2}\frac{|\vec{p}_1|}{m^2_{\Xi^{*0}}}\bar{|{\cal{M}}|^2}d\Omega,\\
d\Gamma(\Xi(1620)^{0}&\to\gamma\Xi\pi,\gamma\bar{K}\Lambda)=\frac{1}{2J+1}\frac{1}{(2\pi)^5}\frac{1}{16m^2}\nonumber\\
                                  &\times{}\bar{|{\cal{M}}|^2}|\vec{p}^{*}_3||\vec{p}_2|dm_{13}d\Omega^{*}_{p_3}d\Omega_{p_2},
\end{align}
where  $J$ is the total angular momentum of the $\Xi(1620)$,  $|\vec{p}_1|$ is the three-momenta of the decay products in the center
of mass frame, the overline indicates the sum over the polarization vectors of the final hadrons.  The ($\vec{p}^{*}_3,\Omega^{*}_{p_3}$)
is the momentum and angle of the particle $\pi$ in the rest frame of $\pi$ and $\Xi$, and $\Omega_{p_2}$ is the angle of the photon in
the rest frame of the decaying particle.  The $m_{13}$ is the invariant mass for $\pi$ and $\Xi$ and $m_{\pi}+m_{\Xi}\leq{}m_{13}\leq{}m$.

\section{NUMERICAL RESULTS AND DISCUSSIONS}\label{sec:5}
\begin{figure}[htbp]
\centering
\includegraphics[bb=00 00 780 540, clip, scale=0.25]{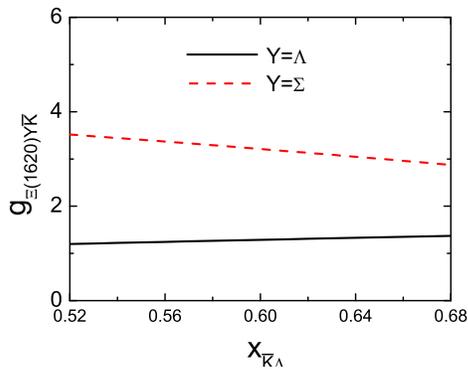}
\caption{The $x_{\bar{K}\Lambda}$ dependence of $g_{\Xi^{*}(1620)\bar{K}\Lambda}$
and $g_{\Xi^{*}(1620)\bar{K}\Sigma}$. }
\label{fig:coup}
\end{figure}
To compute the radiative decay widths of the considered processes, the coupling constants of $\Xi^{*}(1620)$ to his
components could be estimated by the compositeness conditions given by Eq.~\ref{eqn3}.   Regarding the $\Xi^{*}(1620)$
as $S$-wave loosely $\bar{K}\Lambda$-$\bar{K}\Sigma$ hadronic molecule, the coupling constants $g_{\Xi^{*}(1620)\bar{K}\Lambda}$
and $g_{\Xi^{*}(1620)\bar{K}\Sigma}$ dependent on the parameter $x_{\bar{K}\Lambda}$ are plotted in Fig.~\ref{fig:coup}.
The parameter $x_{\bar{K}\Lambda}$ was constrained as $x_{\bar{K}\Lambda}=0.52-0.68$ by comparing the sum of the partial
decay modes of $\Xi^{*}(1620)$ with the total width in our previous paper~\cite{Huang:2020taj}.  It means that about $52\%-68\%$
of the total width comes from the $\bar{K}\Lambda$ channel,  while the rest is provided by the $\bar{K}\Sigma$ channel.
We find that the coupling constant $g_{\Xi^{*}(1620)\bar{K}\Lambda}$ monotonously increase with the increasing of the parameter
$x_{\bar{K}\Lambda}$ in our consider $x_{\bar{K}\Lambda}$ range, while the coupling constant $g_{\Xi^{*}(1620)\bar{K}\Sigma}$
decline decreases with increasing $x_{\bar{K}\Lambda}$.  The opposite trend can be easily understood, as the coupling constants
$g_{\Xi^{*}(1620)\bar{K}\Lambda}$ and $g_{\Xi^{*}(1620)\bar{K}\Sigma}$ are directly proportional to the corresponding
molecular compositions~\cite{Dong:2009uf}.

\begin{figure}[htbp]
\centering
\includegraphics[scale=0.40]{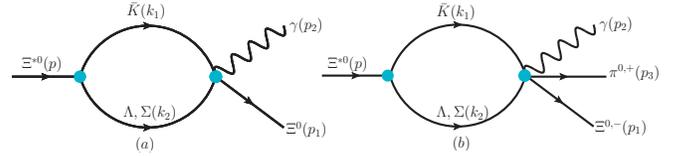}
\caption{Contact diagram for $\Xi^{*0}\to\Xi^{0}\gamma$, $\Xi^{*0}\to\Xi^{0}\pi^0\gamma$, and $\Xi^{*0}\to\pi^{+}\Xi^{-}\gamma$.
We also show definitions of the kinematics ($p_1,p_2,p_3,k_1,k_2$, and $p$) used in the calculation.}
\label{fig:contact}
\end{figure}

With obtained above coupling constants, the radiative decay width of the $\Xi(1620)^0$ into $\Xi^0\gamma$, $\Xi\gamma\pi$,
and $\gamma\bar{K}\Lambda$ that are shown in Fig.~\ref{fety} and Fig.~\ref{fig:example1} can be calculated straightforwardly.
However, the amplitudes of the $\Xi(1620)^0\to\gamma\Xi^0$ and $\Xi(1620)^0\to\gamma\Xi\pi$ cannot satisfy the gauge invariance
of the photon field.  To ensure the gauge invariance of the total amplitudes, the contact diagram must be included as well.
The corresponding Feynman diagrams are shown in Fig.~\ref{fig:contact}.
For the present calculation, we adopt the following form to satisfy $p_2^{\mu}{\cal{M}}^{Total}_{\mu}(\equiv{{\cal{M}}}^{T}+{\cal{M}}_{com})=0$
\begin{align}
&{\cal{M}}^a_{com}(\Xi^{*}\to\gamma\Xi^0)=\frac{D+F}{\sqrt{2}f}eC_{Y}g_{\Xi^{*}\Sigma^{+}K^{-}}\int_0^{\infty}d\alpha\int_0^{\infty}d\eta\nonumber\\
                 &\times\int_0^{\infty}d\zeta\frac{1}{16\pi^2y^2\beta^2}\bar{u}(p_1)({\cal{C}}_1^{\mu}{\cal{T}}_1+{\cal{C}}_2^{\mu}{\cal{T}}_2)\gamma_5u(p)\epsilon^{*}_{\mu}(p_2),\\
\begin{split}
&{\cal{M}}^{b}_{com}(\Xi^{*0}\to\pi^0\Xi^{0}\gamma,\pi^{+}\Xi^{-}\gamma)=-i^3\frac{eC_{Y}g_{\Xi^{*}K^{-}\Sigma^{+}}}{4f^2}\left \{
\begin{array}{ll}
   \frac{1}{\sqrt{2}}\\
    1 \\
\end{array}
\right.
\end{split}
\nonumber\\
&\times\int_0^{\infty}d\alpha\int_0^{\infty}d\eta\int_0^{\infty}d\zeta\frac{1}{16\pi^2y^2\beta^2}\nonumber\\
&\times\bar{u}(p_1)({\cal{D}}_1^{\mu}{\cal{T}}_1+{\cal{D}}_2^{\mu}{\cal{T}}_2)u(p)\epsilon^{*}_{\mu}(p_2).
\end{align}
Where
\begin{align}
{\cal{T}}_1&=\exp\{-\frac{1}{\beta^2}[\alpha(-p_1^2+m^2_{K^{-}})+\eta{}m_{\Sigma^{+}}^2+\zeta(-p_2^2+m^2_{\Sigma^{+}})\nonumber\\
           &-(p_1\omega_{\Sigma^{+}}-p_2\omega_{K^{-}})^2+\frac{1}{4z}({\cal{H}}_2p_2-{\cal{H}}_1p_1)^2]\},\\
{\cal{T}}_2&=\exp\{-\frac{1}{\beta^2}[\alpha(-p_2^2+m^2_{K^{-}})+\eta{}m_{K^{-}}^2+\zeta(-p_1^2+m^2_{\Sigma^{+}})\nonumber\\
           &-(p_2\omega_{\Sigma^{+}}-p_1\omega_{K^{-}})^2+\frac{1}{4z}({\cal{H}}_2p_1-{\cal{H}}_1p_2)^2]\},\\
{\cal{C}}_1^{\mu}&=-\frac{m_1^2{\cal{H}}_1^3}{8y^3}(2p_1^{\mu}-m_1\gamma^{\mu})-\frac{m_1^3{\cal{H}}_1^2}{4y^2}\gamma^{\mu}+\frac{m_1^2{\cal{H}}_1^2}{2y^2}p_1^{\mu}\nonumber\\
                 &-\frac{m_1m^2_{\Sigma^{+}}{\cal{H}}_1}{2y}\gamma^{\mu}+m_1m^2_{\Sigma^{+}}\gamma^{\mu}+\frac{m_1m_{\Sigma^{+}}{\cal{H}}^2_1}{2y^2}p_1^{\mu}\nonumber
\end{align}
\begin{align}
                 &+(\frac{(m+m_1)m_{\Sigma^{+}}{\cal{H}}_1{\cal{H}}_2}{2y^2}-\frac{m_1m_{\Sigma^{+}}{\cal{H}}_1}{y})p_1^{\mu}-\frac{m_{\Sigma^{+}}-m_1}{y}\gamma^{\mu}\nonumber\\
                 &+\frac{{\cal{H}}_1^2{\cal{H}}_2}{4y^3}p_1\cdot{}p_2(2p_1^{\mu}-m_1\gamma^{\mu})+\frac{m_1{\cal{H}}_1{\cal{H}}_2}{2y^2}p_1\cdot{}p_2\gamma^{\mu}\nonumber\\
                 &+\frac{(m+m_1)m_1{\cal{H}}_1{\cal{H}}_2}{2y^2}p_1^{\mu}+\frac{3{\cal{H}}_1}{2y^2}(2p_1^{\mu}-m_1\gamma^{\mu}),\\
{\cal{C}}_2^{\mu}&=-\frac{m_1^2{\cal{H}}_2^3}{4y^3}p_1^{\mu}-\frac{m_1^2{\cal{H}}_2^2}{4y^2}p_1^{\mu}-\frac{(m+m_1)m_{\Sigma^{+}}{\cal{H}}_1{\cal{H}}_2}{2y^2}p_1^{\mu}\nonumber\\
                 &-\frac{(m+m_1)m_{\Sigma^{+}}{\cal{H}}_1}{2y}p_1^{\mu}-\frac{m_1m_{\Sigma^{+}}{\cal{H}}^2_2}{2y^2}p_1^{\mu}-\frac{m_1m_{\Sigma^{+}}{\cal{H}}_2}{2y}p_1^{\mu}\nonumber\\
                 &+\frac{m_{\Sigma^{+}}}{y}\gamma^{\mu}+\frac{{\cal{H}}_1{\cal{H}}^2_2}{2y^3}p_1\cdot{}p_2p_1^{\mu}+\frac{{\cal{H}}_1{\cal{H}}_2}{2y^2}p_1\cdot{}p_2p_1^{\mu}\nonumber\\
                 &-\frac{m_1(m+m_1){\cal{H}}^2_2}{2y^2}p_1^{\mu}-\frac{m_1(m+m_1){\cal{H}}_2}{2y}p_1^{\mu}\nonumber\\
                 &+\frac{{\cal{H}}_2}{2y^2}(2p_1^{\mu}-m_1\gamma^{\mu})+\frac{m_1{\cal{H}}_2}{2y^2}\gamma^{\mu}+\frac{2{\cal{H}}_2}{y^2}p_1^{\mu}+\frac{2}{y}p_1^{\mu},\\
{\cal{D}}_1^{\mu}&=(-\frac{m^2_{\Sigma^{+}}{\cal{H}}_1}{2y}+m^2_{\Sigma^{+}})(2p^{\mu}-m\gamma^{\mu})-\frac{m_{\Sigma^{+}}{\cal{H}}_1^2}{2y^2}(m-p\!\!\!/_2)p^{\mu}\nonumber\\
                 &+\frac{m_{\Sigma^{+}}{\cal{H}}_1{\cal{H}}_2}{2y^2}p\!\!\!/_2p^{\mu}+\frac{m_{\Sigma^{+}}{\cal{H}}_1}{y}(m-p\!\!\!/_2)p^{\mu}+\frac{m_{\Sigma^{+}}}{y}\gamma^{\mu}\nonumber\\
                 &-\frac{mm_{13}^2{\cal{H}}^3_1}{8y^3}\gamma^{\mu}+\frac{m(m^2-m^2_{13}){\cal{H}}^2_1{\cal{H}}_2}{8y^3}\gamma^{\mu}-\frac{m^2_{13}{\cal{H}}^2_1}{4y^2}\nonumber\\
                 &\times(2p^{\mu}-m\gamma^{\mu})+\frac{{\cal{H}}^2_1m^2_{13}}{2y^2}p^{\mu}+(\frac{{\cal{H}}_1{\cal{H}}_2(m^2-m^2_{13})}{4y^2}+\frac{1}{y})\nonumber\\
                 &\times(2p^{\mu}-m\gamma^{\mu})-\frac{m{\cal{H}}_1{\cal{H}}_2}{2y^2}p\!\!\!/_2p^{\mu}+\frac{3m{\cal{H}}_1}{2y^2}\gamma^{\mu},\\
{\cal{D}}_2^{\mu}&=-\frac{m_{\Sigma^{+}}{\cal{H}}_1{\cal{H}}_2}{2y^2}p\!\!\!/_2p^{\mu}+\frac{m_{\Sigma^{+}}{\cal{H}}^2_2}{2y^2}(m-p\!\!\!/_2)p^{\mu}-\frac{m_{\Sigma^{+}}}{y}\gamma^{\mu}\nonumber\\
                 &+\frac{(m^2-m_{13}^2){\cal{H}}_1{\cal{H}}_2^2}{4y^3}p^{\mu}-\frac{m{\cal{H}}_1{\cal{H}}_2}{2y^2}p\!\!\!/_2p^{\mu}-\frac{m^2_{13}{\cal{H}}_2^3}{4y^3}p^{\mu}\nonumber\\
                 &+\frac{m^2_{13}{\cal{H}}_2^2}{2y^2}p^{\mu}+\frac{3{\cal{H}}_2}{y^2}p^{\mu}-\frac{m}{y}\gamma^{\mu}.
\end{align}
with $y=1+\alpha+\eta+\zeta$,${\cal{H}}_2=2(\zeta+\omega_{K^{-}})$,${\cal{H}}_1=2(\alpha+\omega_{\Sigma^{+}})$, and $m^2_{13}=(p_1+p_3)^2$.
$m$ and $m_1$ are the masses of the $\Xi^{*}$ and $\Xi$, respectively.

\begin{figure}[htbp]
\centering
\includegraphics[bb=00 00 780 540, clip, scale=0.25]{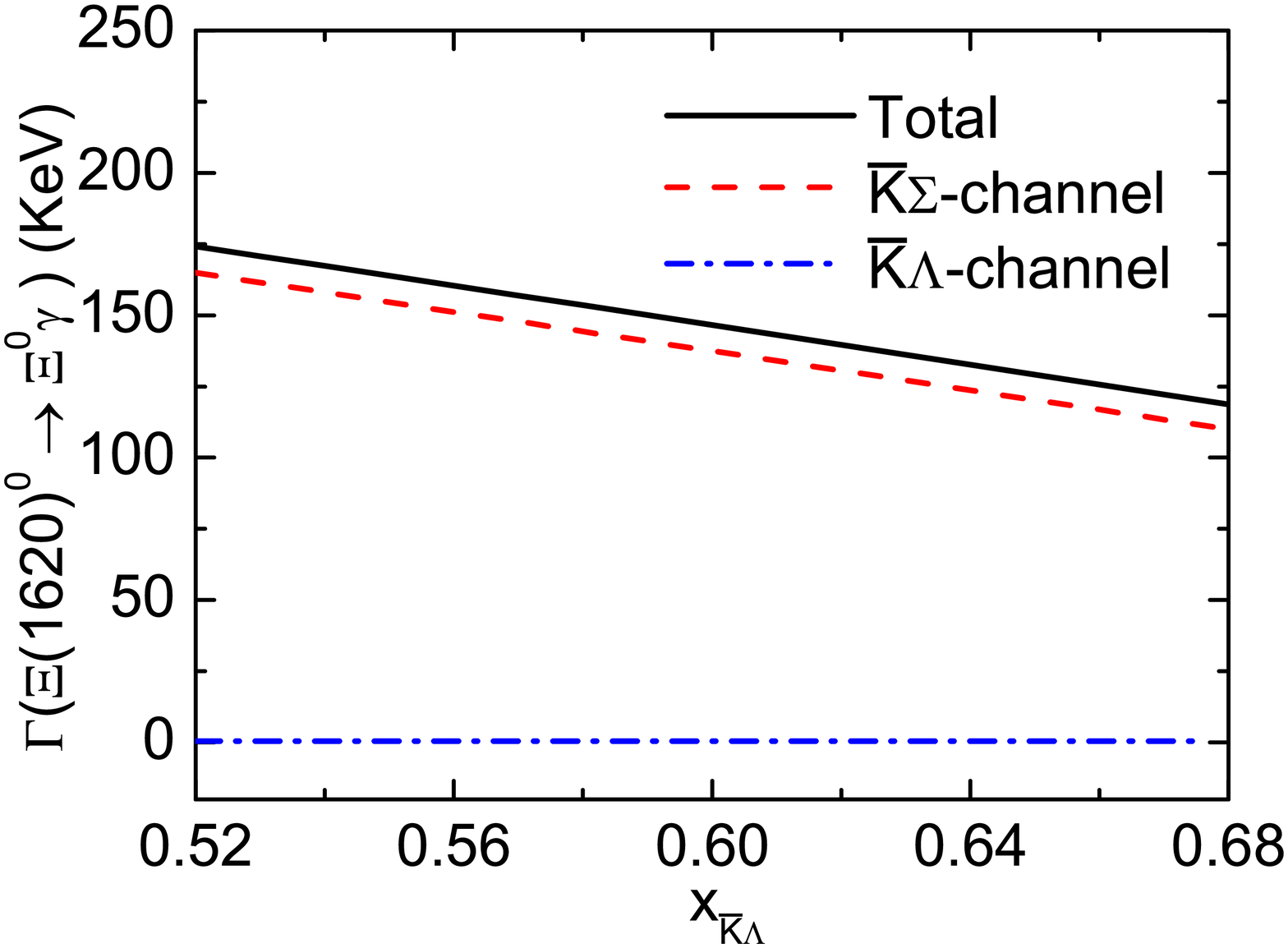}
\caption{Decomposed contributions to the decay width of the $\Xi(1620)^0$ into $\Xi^0\gamma$. }
\label{fig:width1}
\end{figure}
With obtained the total amplitude,  the radiative decay width of the $\Xi(1620)^0$ into $\Xi^0\gamma$ can be
calculated straightforwardly.  The dependence of the corresponding radiative decay width on $x_{\bar{K}\Lambda}$ is
given in Fig.~\ref{fig:width1}.  The $x_{\bar{K}\Lambda}$ value within a reasonable range from 0.52 to 0.68, the
radiative decay width of the $\Xi(1620)^0\to\gamma\Xi^0$ contribution from the $\bar{K}\Sigma$ channel monotonously
decreases.   However, it increases for the $\bar{K}\Lambda$ channel contribution to the $\Xi(1620)^0\to\gamma\Xi^0$.
Moreover,  the $\bar{K}\Sigma$ component provides the dominant contribution to the partial decay width of the
$\gamma\Xi^0$ two-body channel.  The $\bar{K}\Lambda$ contribution to the $\gamma\Xi^0$ two-body channel is very small.
This is different from our results in Ref.~\cite{Huang:2020taj} that the $\bar{K}\Lambda$ component provides the
dominant contribution to the strong decay width of the $\Xi(1620)$.  A possible explanation for these may be that
the interaction between the $\Sigma$ baryon and photon is stronger than $\gamma{}-\Lambda$ interaction because since
the $\Sigma^0$ decays completely to the final state containing $\Lambda$ baryon and $\gamma$~\cite{Zyla:2020zbs}.

Fig.~\ref{fig:width1} also tell us that the total radiative decay width decrease for the $\Xi(1620)^0\to\gamma\Xi^0$ when the
$x_{\bar{K}\Lambda}$ is changed from 0.52 to 0.68.  We also find that the interference between the $\bar{K}\Sigma$ channel and
$\bar{K}\Lambda$ channel is quite small, leading to a total decay width of the $\Xi(1620)^0\to\gamma\Xi^0$ mainly contribution
from the $\bar{K}\Sigma$ channel.   This does not alter the conclusion that the $\bar{K}\Lambda$ channel strongly couples
to the $\bar{K}\Sigma$ channel~\cite{Ramos:2002xh,Huang:2020taj}.   The main reason for this is that the radiative decay widths
are often in the keV regime and are far less than their strong counterparts.   Indeed, the total radiative decay width for the
$\Xi(1620)^0\to\gamma\Xi^0$ is predicted to be $118.76-174.21$ KeV, which is far less than the total decay width is predicted
to be about 50.39-68.79 MeV~\cite{Huang:2020taj}.
\begin{figure}[htbp]
\centering
\includegraphics[bb=00 80 600 780, clip, scale=0.25]{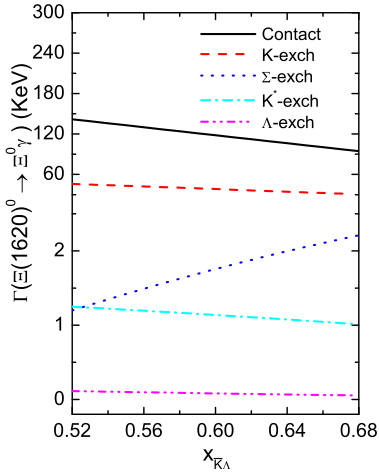}
\caption{(color online) The partial decay widths from $K^{-}$(red dash line), $\bar{K}^{*}$(cyan dash dot line), $\Sigma$ (blue dot line),
$\Lambda$(magenta dash dot dot line), and the remainder is the contact term exchange contribution for the $\Xi(1620)^0\to\gamma\Xi^0$ as a
function of the parameter $x_{\bar{K}\Lambda}$.}
\label{fig:didi}
\end{figure}

The individual contributions of $K^{-}$, $\bar{K}^{*}$, $\Sigma$, $\Lambda$ exchanges, and contact term for the reaction $\Xi(1620)^0\to\gamma\Xi^0$
are shown in Fig.~\ref{fig:didi}.  The amplitudes corresponding to the $K^{-}$-exchange and $\Sigma^{+}$-exchange are not gauge invariant,
while the rest is  gauge invariant.  One can see that the contract term and $K^{-}$-exchange provide a dominant contribution to the total
decay width, and is at last sixty orders of magnitude bigger than those of the amplitudes corresponding to the $\bar{K}^{*}$, $\Sigma$,
and $\Lambda$ exchanges for the $x_{\bar{K}\Lambda}$ range studied.

\begin{figure}[htbp]
\centering
\includegraphics[bb=00 25 500 480, clip, scale=0.35]{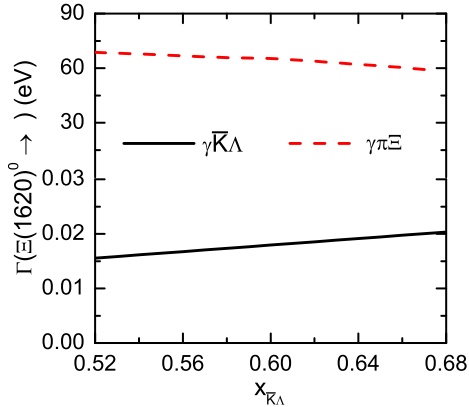}
\caption{(color online) The partial decay widths of the $\Xi(1620)^0\to\gamma\Xi\pi$ and $\Xi(1620)^0\to\gamma\bar{K}\Lambda$.}
\label{fig:santi}
\end{figure}
Now we turn to the three-body radiative decays $\Xi(1620)^0\to\gamma\pi\Xi$ and $\Xi(1620)^0\to\gamma\bar{K}\Lambda$.  The decay widths
with varying $x_{\bar{K}\Lambda}$ from 0.52 to 0.68 for such two transitions are presented in Fig.~\ref{fig:santi}.   Since the phase
space is small compared with the two-body radiative decay channel $\Xi(1620)^0\to\gamma\Xi^0$,  the decay width should be the smallest
for the $\Xi(1620)^0\to\gamma\bar{K}\Lambda$ channel, should be intermediate for the $\Xi(1620)^0\to\gamma\pi\Xi$ channel, and should be
biggest for the $\Xi(1620)^0\to\gamma\Xi^0$ channel.   Indeed, our study shows that the partial width of $\Xi(1620)^0\to\gamma\bar{K}\Lambda$
is rather small, weakly increasing with the $x_{\bar{K}\Lambda}$ increasing.  In particular, the partial width varies from 0.016 to
0.020 eV in the $x_{\bar{K}\Lambda}$ range studied.   However, the partial width of the $\Xi(1620)^0\to\gamma\Xi\pi$ decreases
with the increase of $x_{\bar{K}\Lambda}$ and the partial width of the $\Xi(1620)^0\to\gamma\Xi\pi$ is estimated to be $68.75-58.19$ eV.

\section{SUMMARY}
We studied the two-body and three-body radiative decays of the $\Xi(1620)$ state assuming that it is a bound state of
$\bar{K}\Lambda$-$\bar{K}\Sigma$.   The coupling of $\Xi(1620)$ to his components are fixed by the Weinberg compositeness
condition.   The radiative decays for the $\Xi(1620)^0\to\gamma\Xi^0$ and $\Xi(1620)^0\to\gamma\pi\Xi$ are via triangle
diagrams with exchanges of a pseudoscalar meson $K$, vector meson $K^{*}$, and baryon $\Sigma$ and $\Lambda$.  The three
body decay for the $\Xi(1620)^0\to\gamma\bar{K}\Lambda$ happen at tree level.  In the relevant parameter region, the partial
widths are evaluated as
\begin{align}
&\Gamma(\Xi(1620)^0\to\gamma\Xi^0)=118.76-174.21 ~\rm{KeV},\nonumber\\
&\Gamma(\Xi(1620)^0\to\gamma\Xi\pi)=58.19-68.75  ~\rm{eV},\\
&\Gamma(\Xi(1620)^0\to\gamma\bar{K}\Lambda)=0.016-0.020 ~\rm{eV}.\nonumber
\end{align}
Future experimental measurements of these processes can be useful to test the molecule interpretations of the $\Xi(1620)$.
Based on the current integrated luminosity and our estimations,  facilities  such  as  the  LHCb  might  be  able  to
detect radiative decays of the baryon in the keV regime.  This research can also be performed in the forthcoming Belle II
experiment.

\begin{acknowledgments}
This work was supported by the Science and Technology Research Program of Chongqing Municipal Education
Commission (Grant No. KJQN201800510), the Opened Fund of the State Key Laboratory on Integrated Optoelectronics
(GrantNo. IOSKL2017KF19).  Yin Huang want to thanks the support from the Development and Exchange Platform for the
Theoretic Physics of Southwest Jiaotong University under Grants No.11947404 and No.12047576, the Fundamental
Research Funds for the Central Universities(Grant No. 2682020CX70), and the National Natural Science Foundation
of China under Grant No.12005177.
\end{acknowledgments}

\end{document}